\documentclass[twoside,twocolumn,9pt]{article}
\usepackage{extsizes}
\usepackage[super,sort&compress,comma]{natbib} 
\usepackage[version=3]{mhchem}
\usepackage[left=1.5cm, right=1.5cm, top=1.785cm, bottom=2.0cm]{geometry}
\usepackage{balance}
\usepackage{mathptmx}
\usepackage{sectsty}
\usepackage{graphicx} 
\usepackage{lastpage}
\usepackage[format=plain,justification=justified,singlelinecheck=false,font={stretch=1.125,small,sf},labelfont=bf,labelsep=space]{caption}
\usepackage{float}
\usepackage{fancyhdr}
\usepackage{fnpos}
\usepackage[english]{babel}
\addto{\captionsenglish}{%
  
}
\usepackage{array}
\usepackage{droidsans}
\usepackage{charter}
\usepackage[T1]{fontenc}
\usepackage[usenames,dvipsnames]{xcolor}
\usepackage{setspace}
\usepackage[compact]{titlesec}
\usepackage{hyperref}
\usepackage{epstopdf}%This line makes .eps figures into .pdf - please comment out if not required.

\usepackage{pgfplots}% Added by MAM

\definecolor{cream}{RGB}{222,217,201}

\begin{document}

\pagestyle{fancy}
\thispagestyle{plain}
\fancypagestyle{plain}{
%%%HEADER%%%
\renewcommand{\headrulewidth}{0pt}
}
%%%END OF HEADER%%%

%%%PAGE SETUP - Please do not change any commands within this section%%%
\makeFNbottom
\makeatletter
\renewcommand\LARGE{\@setfontsize\LARGE{15pt}{17}}
\renewcommand\Large{\@setfontsize\Large{12pt}{14}}
\renewcommand\large{\@setfontsize\large{10pt}{12}}
\renewcommand\footnotesize{\@setfontsize\footnotesize{7pt}{10}}
\makeatother

\renewcommand{\thefootnote}{\fnsymbol{footnote}}
\renewcommand\footnoterule{\vspace*{1pt}% 
\color{cream}\hrule width 3.5in height 0.4pt \color{black}\vspace*{5pt}} 
\setcounter{secnumdepth}{5}

\makeatletter 
\renewcommand\@biblabel[1]{#1}            
\renewcommand\@makefntext[1]% 
{\noindent\makebox[0pt][r]{\@thefnmark\,}#1}
\makeatother 
\renewcommand{\figurename}{\small{Fig.}~}
\sectionfont{\sffamily\Large}
\subsectionfont{\normalsize}
\subsubsectionfont{\bf}
\setstretch{1.125} %In particular, please do not alter this line.
\setlength{\skip\footins}{0.8cm}
\setlength{\footnotesep}{0.25cm}
\setlength{\jot}{10pt}
\titlespacing*{\section}{0pt}{4pt}{4pt}
\titlespacing*{\subsection}{0pt}{15pt}{1pt}
%%%END OF PAGE SETUP%%%

%%%FOOTER%%%
\fancyfoot{}
\fancyfoot[LO,RE]{\vspace{-7.1pt}\includegraphics[height=9pt]{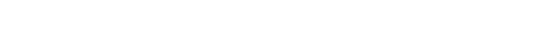}}
\fancyfoot[CO]{\vspace{-7.1pt}\hspace{13.2cm}\includegraphics{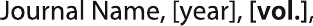}}
\fancyfoot[CE]{\vspace{-7.2pt}\hspace{-14.2cm}\includegraphics{head_foot/RF}}
\fancyfoot[RO]{\footnotesize{\sffamily{1--\pageref{LastPage} ~\textbar  \hspace{2pt}\thepage}}}
\fancyfoot[LE]{\footnotesize{\sffamily{\thepage~\textbar\hspace{3.45cm} 1--\pageref{LastPage}}}}
\fancyhead{}
\renewcommand{\headrulewidth}{0pt} 
\renewcommand{\footrulewidth}{0pt}
\setlength{\arrayrulewidth}{1pt}
\setlength{\columnsep}{6.5mm}
\setlength\bibsep{1pt}
%%%END OF FOOTER%%%

%%%FIGURE SETUP - please do not change any commands within this section%%%
\makeatletter 
\newlength{\figrulesep} 
\setlength{\figrulesep}{0.5\textfloatsep} 

\newcommand{\topfigrule}{\vspace*{-1pt}% 
\noindent{\color{cream}\rule[-\figrulesep]{\columnwidth}{1.5pt}} }

\newcommand{\botfigrule}{\vspace*{-2pt}% 
\noindent{\color{cream}\rule[\figrulesep]{\columnwidth}{1.5pt}} }

\newcommand{\dblfigrule}{\vspace*{-1pt}% 
\noindent{\color{cream}\rule[-\figrulesep]{\textwidth}{1.5pt}} }

\makeatother
%%%END OF FIGURE SETUP%%%

%%%TITLE, AUTHORS AND ABSTRACT%%%
\twocolumn[
  \begin{@twocolumnfalse}
{\includegraphics[height=30pt]{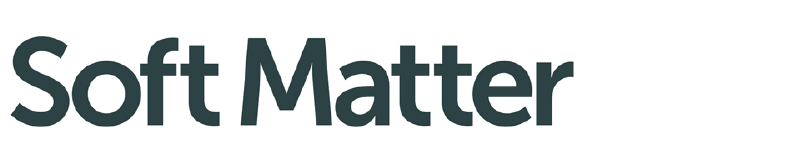}\hfill\raisebox{0pt}[0pt][0pt]{\includegraphics[height=55pt]{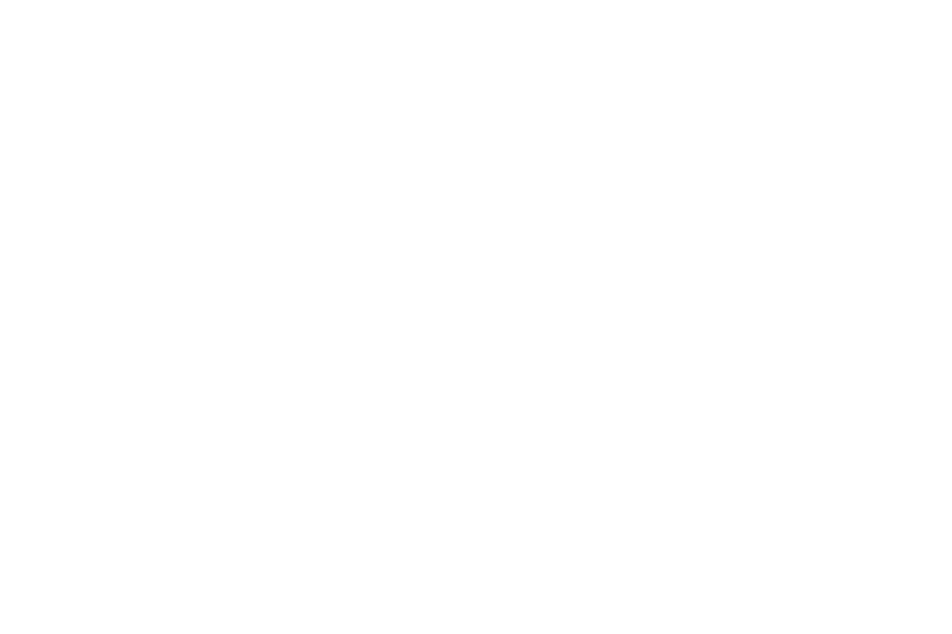}}\\[1ex]
\includegraphics[width=18.5cm]{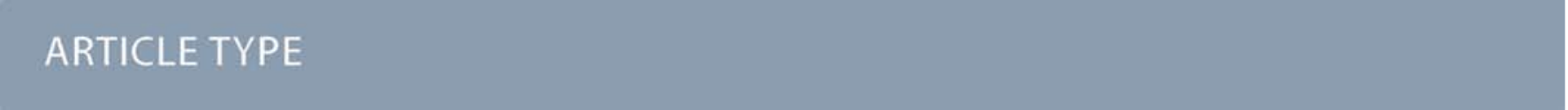}}\par
\vspace{1em}
\sffamily
\begin{tabular}{m{4.5cm} p{13.5cm} }

\includegraphics{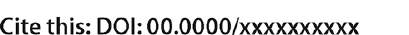} & \noindent\LARGE{\textbf{Phase transitions on non-uniformly curved surfaces: Coupling between phase and location$^\dag$}} \\
\vspace{0.3cm} & \vspace{0.3cm} \\

 & \noindent\large{Jack O.~Law,\textit{$^{a}$} Jacob M.~Dean,\textit{$^{b\ddag}$}, Mark A.~Miller,$^{{\textit{b}\text{\textsection}}}$ and Halim Kusumaatmaja\textit{$^{{a}\P}$}} \\%Author names go here instead of "Full name", etc.

\includegraphics{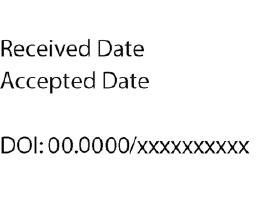} & \noindent\normalsize{
For particles confined to two dimensions, any curvature of the surface affects
the structural, kinetic and thermodynamic properties of the system.  If the curvature
is non-uniform, an even richer range of behaviours can emerge.  Using a combination of
bespoke Monte Carlo, molecular dynamics and basin-hopping methods, we show that the stable
states of attractive colloids confined to non-uniformly curved surfaces are
distinguished not only by the phase of matter but also by their location
on the surface.  Consequently, the transitions between these states involve cooperative
migration of the entire colloidal assembly.  We
demonstrate these phenomena on toroidal and sinusoidal surfaces for model colloids
with different ranges of interactions as described by the Morse potential.
In all cases, the behaviour can be rationalised in terms of three universal
considerations: cluster perimeter, stress, and the packing of next-nearest neighbours.
} \\

\end{tabular}

 \end{@twocolumnfalse} \vspace{0.6cm}

  ]
%%%END OF TITLE, AUTHORS AND ABSTRACT%%%

%%%FONT SETUP - please do not change any commands within this section
\renewcommand*\rmdefault{bch}\normalfont\upshape
\rmfamily
\section*{}
\vspace{-1cm}

%%%FOOTNOTES%%%

\footnotetext{\dag~Electronic Supplementary Information (ESI) available: [RSC DOI].  For simulation code and archives of raw data see DOI:10.15128/r1wh246s13w.}
\footnotetext{\textit{$^{a}$Department of Physics, Durham University, South Road, Durham DH1 3LE, United Kingdom.}}
\footnotetext{\textit{$^{b}$Department of Chemistry, Durham University, South Road, Durham DH1 3LE, United Kingdom.}}

\footnotetext{{\ddag}Present address: Department of Chemistry, University of Bath, Claverton Down, Bath BA2 7AY, United Kingdom.}

\footnotetext{{\textsection}E-mail: m.a.miller@durham.ac.uk}
\footnotetext{{\P}E-mail: halim.kusumaatmaja@durham.ac.uk}

%%%END OF FOOTNOTES%%%

%%%MAIN TEXT%%%%

\section{Introduction}

Two-dimensional systems in which particles are confined to surfaces of non-uniform curvature
abound in nature.  For example, non-uniformly curved regions in cellular membranes are
necessary for many key biological processes, including the sensing and trafficking properties
of organelles such as the Golgi apparatus.\cite{Vahid2017A}  Non-uniform curvature is also
found in the capsid of the torovirus, which infects agricultural animals,\cite{Mark2008A}
and in the cubic phases of lipids commonly used in the formulation and food
industries.\cite{Paillusson2016A}  Furthermore, it is becoming increasingly realistic
to engineer artificial surfaces with specified curvature.  Techniques include a rotating
cuvette,\cite{Ellis2017A} lithography,\cite{Liu2013A} suspending a liquid surface from a
post,\cite{Stebe2017A} and 3D printing.\cite{Talha2018A}  These surfaces can play host
to a variety of two-dimensional systems, including membrane proteins,\cite{Vahid2017A}
stress fibres,\cite{Stebe2018A} active and passive liquid
crystals,\cite{Keber2014A,Nieves2011A} capsid protein shells\cite{Paquay17a,Mark2008A}
and colloidal particles adsorbed onto a surface
by depletion forces or tethered to one with DNA.\cite{Meng2014A,Joshi2016A}
Additionally, it has been shown that the dimple patterns in buckled curved
elastic bilayers form a crystal structure with similar
properties to colloidal crystals on a curved surface.\cite{Reis2016A}
\par
The practical importance of these systems, as well as their rich and novel behaviour, mark
them out for considerable scientific interest.  So far, most studies on the effects of
curvature have focused on the case of constant curvature, such as spheres.  Even in this
simplest scenario, a wide range of phenomena are observed which are absent on flat surfaces,
including the presence of defects and branching in the ground-state
crystals,\cite{Meng2014A,Bausch2003A,Yao2017A} and of modified nucleation
pathways.\cite{Gomez2015A,Law2018A}  These studies have been successful in describing
natural phenomena such as the structure and formation of virus capsids,\cite{Paquay17a}
and the packing of particles on a Pickering emulsion droplet.\cite{Bausch2003A}
\par
An even richer picture emerges for surfaces with varying curvature, where the symmetry of the surface is
broken.  Nucleating phases form preferentially in certain regions due to the underlying curvature, and
the free energy profiles may include additional metastable
minima.\cite{Gomez2015A,Marenduzzo2012A}
Crystal defects feel a local potential that arises purely from the underlying
curvature.\cite{Vitelli2006A,Reis2016A}  Topological defects also have preferential locations that
depend on their type, as shown for nematic liquid crystals\cite{Ellis2017A} and colloidal
particles.\cite{Kusumaatmaja2013A,Irvine2010A}
\par
To date, however, there is still no complete picture of how non-uniform curvature affects the
thermodynamics of two-dimensional systems.  In this article we consider clusters of attractive
colloidal particles confined to non-uniformly curved surfaces.  We identify three effects --- relating
to cluster perimeter, local stress and the energetics of packing --- and their impacts on the
structure and phase behaviour.  An interesting consequence of these considerations is that the
stable phases (gas, liquid, crystal) are located in different regions of the surface, and phase
transitions involve global translation of the cluster.  While curvature sensing has previously
been reported, in most cases it is due to the architecture or anisotropy of the particle or the
molecule involved.\cite{McMahon2005A,Bruno2011A}  In contrast, here the particles are spherical
and curvature sensing is instead due to a cooperative effect.  Moreover, the effect is
phase-dependent: it acts differently for liquid and crystal phases.  It is even possible for
new states to emerge, producing states with the same phase of matter but in different locations.
\par
We begin with a brief description of our simulation methods, before presenting a phase diagram for
particles on a torus in the canonical ensemble and studying the transitions between distinguishable
states.  Finally, we demonstrate the generality of the effects by presenting similar phenomena on a
sinusoidal surface.

\section{Model and Methods}

\subsection{Potential}

\begin{figure}[t]
\centering
\includegraphics[width=90mm]{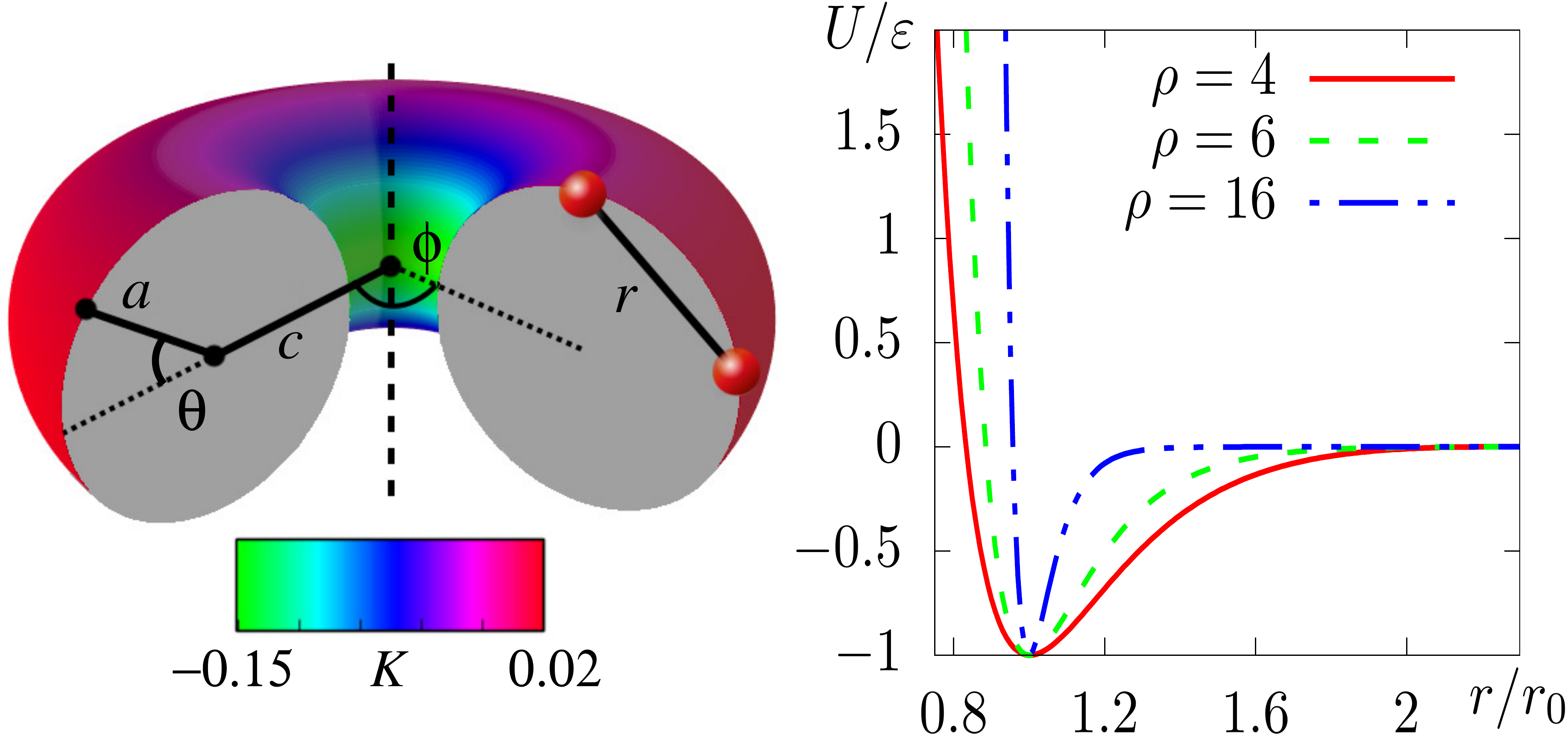}
\caption{\label{fig:torusPlan} Left: Toroidal surface, coloured by the local Gaussian curvature
$K$ in units of $1/r_0^2$. The two size parameters $a$ and $c$ are labelled on the
cross-section.  The angles $\theta$ and $\phi$ specify the location of particles on
the surface.  Also depicted are two particles and the line through which the Morse potential acts.
Note that the potential depends on the Euclidean distance marked $r$ in three-dimensional
space, rather than on the geodesic separation of the particles on the surface of the torus.
Right: The truncated, shifted, smoothed and scaled Morse potential for some representative values
of $\rho$.}
\end{figure}

We model the interactions between the particles with a truncated, shifted and smoothed (tss)
Morse potential, rescaled to restore its well-depth $\varepsilon$ from before the shift. This
potential has an adjustable range parameter $\rho$, which is important because the behaviour
of crystals of attractive particles on curved surfaces is known to depend on the softness of
their interactions.\cite{Meng2014A,Paquay17a}  The pair potential energy is given by 
\begin{eqnarray}
& U\left(r\right) = -\varepsilon U_{\text{tss}}\left(r\right)/U_{\text{tss}}\left(r_0\right), \nonumber \\
& U_{\text{tss}}\left(r\right) = \left[U_{\text{M}}(r)-U_{\text{M}}(r_{\text{c}})-(r-r_{\text{c}})
\displaystyle{\frac{dU_{\text{M}}}{dr}\bigg|_{r_{\text{c}}}}\right]\Theta(r_{\text{c}}-r), \label{eq:tss} \\
& U_{\text{M}}\left(r\right) = \varepsilon\operatorname{e}^{-\rho\left(r-r_0\right)}
\left(\operatorname{e}^{-\rho\left(r-r_0\right)}-2\right), \nonumber
\end{eqnarray}
where $r$ is the separation of the two particles and $r_0$ is the equilibrium pair
separation.  These distances are measured along the Euclidean line that
joins the particles in three-dimensional space, rather than along the geodesic
on the two-dimensional surface, as shown in Fig.~\ref{fig:torusPlan}.  The
Heaviside step function $\Theta$ in Eq.~\ref{eq:tss} truncates the potential at a distance
$r_{\text{c}}$.  For this work we have chosen
to set $r_{\text{c}}/r_0=2.23$, in line with comparable
investigations.\cite{Paquay17a,Vest2014A}  The potential is plotted in Fig.~\ref{fig:torusPlan}.

\subsection{Monte Carlo simulations}

Canonical Monte Carlo (MC) simulations are used to survey the thermodynamic properties for a given
number $N$ of particles over a range of temperatures and values of $\rho$.  The particle positions
are specified in a system of two-dimensional curvilinear coordinates that are natural for the surface
in question, such as the toroidal ($\phi$) and
poloidal ($\theta$) angles of the torus in Fig.~\ref{fig:torusPlan}.  Uniformly distributed trial
displacements are made in the curvilinear coordinates of one
particle per MC step, up to a fixed maximum size (chosen to achieve
an acceptance rate of approximately 50\%).  Because of the nonlinearity of the coordinates, the
Metrpolis acceptance criterion must be generalised to the form
\begin{equation}
P_{\rm acc} = \min\left[1,\frac{g_{\rm n}}{g_{\rm o}}\exp
\left(\frac{-\left(U_{\rm n}-U_{\rm o}\right)}{k_{\rm B}T}\right)\right]
\label{eq:acceptance}
\end{equation}
in order to achieve uniform sampling on the surface.
In Eq.~(\ref{eq:acceptance}), $k_{\rm B}$ is Boltzmann's constant and $T$ is the temperature,
$U_{\rm n/o}$ is the total potential energy of the displaced particle at its new/old position,
and $g_{\rm n/o}$ is the square root of the determinant of the metric tensor at the new/old position.
Explicit expressions for the $g$ factors in terms of the relevant curvilinear coordinates will be
stated when the surfaces are introduced.
\par
To simulate a liquid phase covering the entire surface, we will need grand canonical MC,
where $N$ fluctuates in response to an imposed chemical potential $\mu$.  In practice,\cite{Frenkel2002A}
the control parameter is the activity $z(\mu)=A_0 \Lambda^{-2} \exp(\mu/k_{\rm B}T)$, where $A_0$ is the
area of the surface and $\Lambda$ is the thermal de Broglie wavelength.  Working in terms of $z$
avoids the need to specify $\Lambda$.  In order to produce uniformly distributed trial positions for
the particle-insertion moves, we generate random coordinates distributed according to the metric
factor $g$ by rejection sampling.\cite{Martino2010}

\subsection{Molecular dynamics simulations}

Molecular dynamics (MD) simulations will be used to observe transitions between different states on the
phase diagram as a function of time.  In these simulations, three-dimensional Cartesian coordinates are
used and a one-body RATTLE-like constraint is applied to each particle.  The algorithm forces
the particles to remain on the surface and to have velocity vectors that lie in the local tangent
plane of the surface.\cite{Paquay2016B}  We use the implementation by Paquay {\it et al.}\cite{Paquay2016B}
in the LAMMPS package.\cite{Plimpton95a}
\par
For compatibility with the MC simulations, the MD simulations are performed at constant temperature
using a Langevin thermostat.  The damping time is set to $10$ in the natural Morse time units
of $r_0(m/\varepsilon)^{1/2}$, where $m$ is the mass of one particle.

\subsection{Global optimisation}

We use basin-hopping with parallel tempering\cite{Strodel2010} (BHPT) in the GMIN program\cite{GMIN}
to search for ground-state structures, {\it i.e.,} the globally lowest point on the potential energy
landscape of a given system.
\par
The basic basin-hopping algorithm\cite{Wales97a} is a MC simulation on the transformed
potential energy surface
\begin{displaymath}
{\tilde U}({\bf X}) = {\rm lmin}\left[U({\bf X})\right],
\end{displaymath}
in which the energy $\tilde U$ assigned to a configuration ${\bf X}$ is that obtained by performing
a local minimisation (lmin) of the true potential $U$ starting from ${\bf X}$.  Hence, the potential
energy surface is mapped onto a series of plateaux, each corresponding to the basin of attraction
of a mechanically stable structure.  This transformation removes the barriers between directly connected
pairs of minima, thereby facilitating exploration of the surface and identification of the global minimum,
whose energy is not affected.
\par
Basin-hopping calculations can nevertheless become trapped in limited regions of the potential
energy surface, especially if the surface is rough or contains multiple funnels.  As in ordinary
MC simulations, the efficiency of sampling can be enhanced with parallel tempering (also known
as replica exchange).\cite{Lyubartsev92a,Geyer95a}
In BHPT,\cite{Strodel2010} several basin-hopping replicas run in parallel at different temperatures, and
trial moves occasionally attempt to exchange the configurations currently being sampled by two
runs with adjacent temperatures.  An exchange between replicas $i$ and $j$ with reciprocal temperatures
$\beta_i$ and $\beta_j$ is accepted with probability
\begin{displaymath}
P^{\rm exch}_{ij} = {\rm min}\left[1,
\exp\left\{-\big(\beta_i - \beta_j\big)\big({\tilde U}({\bf X}_i) - {\tilde U}({\bf X}_j)\big)\right\}\right].
\end{displaymath}
\par
For BHPT on a torus, a single MC step involves displacing all particles by moving each one
onto a small sphere centred on its original position and then projecting back to the closest point
on the torus.  This procedure does not strictly preserve detailed balance, but this is not important
since BHPT only attempts to locate the global potential energy minimum rather than to sample a
thermodynamic ensemble.  The advantage of these moves is that they produce roughly uniform
displacements at all points on the torus, unlike steps of fixed maximum size in the toroidal
and poloidal angles.

\section{Results}

\subsection{Localised states on a torus\label{sec:states}}

We have chosen a toroidal surface as our primary example
of a surface with varying curvature for the following reasons: it is relatively simple, both
mathematically and conceptually; it has regions of positive and negative Gaussian curvature;
some progress is being made in reproducing it
experimentally;\cite{Ellis2017A,Kim2006A,Haridas2015A}
and related surfaces can be found in nature (for example the torovirus mentioned
above\cite{Mark2008A}).
\par
The toroidal case is governed by three independent length
scales: the major radius $c$ and the minor radius $a$ of the torus, and the
(inverse) range of the Morse
potential $\rho$, all of which can be expressed in relation to the nominal particle diameter
$r_0$ (see Fig.~\ref{fig:torusPlan}).
The metric $g$ factors appearing in Eq.~(\ref{eq:acceptance}) are given by
$g = c + a \cos\left(\theta\right)$, where $\theta$ is defined in
Fig.~\ref{fig:torusPlan}.  We focus on simulations carried out with $N = 300$ particles on a
torus with $a=5r_0$ and $c=7r_0$, which we call a ``5-7 torus''.  This case illustrates all the
general phenomena that we need to discuss.  However, it is natural to ask how the detailed
picture changes upon varying the number of particles or the geometry of the surface.  In
Sec.~S3 of the ESI$^\dag$ we have therefore provided comparisons for smaller and larger values of
$N$ on the same torus, as well as for $N=300$ on the thinner 3.5-10 torus.
\par
Surface curvature can influence the free energy of a cluster on the surface in three ways:
\begin{enumerate}
\item The length of the {\em perimeter} of a cluster of a given area depends on the
underlying curvature, changing the line tension contribution to the free
energy.\cite{Gomez2015A}  In regions of positive Gaussian curvature,
the perimeter is generally smaller than on a surface with zero or negative Gaussian curvature.
\item A hexagonal crystal structure is necessarily distorted in regions of non-zero Gaussian curvature,
leading to a {\em stress} that penalises the free energy.\cite{Meng2014A,Paquay17a}
\item Curvature can make the interactions between next-nearest {\em neighbours}
more favourable.  For example, in a region of large negative Gaussian curvature (a saddle), the area immediately
around a given particle inreases more rapidly with distance on the curved surface than it would on a
plane, allowing next-nearest neighbours to approach more closely.  Furthermore, if the interactions act
through space as they do in our model (rather than geodesically along the curved surface), any region
with a large principal curvature will bring next-nearest neighbours closer in space.  This latter
effect can contribute even if the Gaussian curvature is zero (for example on cylinders and cones).
\end{enumerate}
We will refer to the three contributions by the italicised terms above.
The effects respond to curvature in different ways and they apply to the conventional states
of matter (gas, liquid, solid) to different extents.  The resulting couplings compete with
each other to produce a rich set of stable states that are defined not only by the
phase but also by the location of matter
on the surface.  We will denote the phase by a letter
(G, L, C, X), and the location by the sign of the local Gaussian curvature ($+$, $-$, $0$, $\pm$).
The meanings of the symbols are summarised in Table \ref{tab:names}.

\begin{table}[t]
\small
  \caption{\ Symbols used to denote the phase and location of thermodynamically stable states}
  \label{tab:names}
  \begin{tabular*}{0.48\textwidth}{@{\extracolsep{\fill}}lll}
    \hline
    Symbol & Meaning \\
    \hline
    G & gas phase (covering whole surface)\\
    L & liquid phase \\
    C & condensed phase (intermediate order) \\
    X & crystal phase \\
    \hline
    $+$ & region of positive Gaussian curvature \\
    $-$ & region of negative Gaussian curvature \\
    0 & region of zero Gaussian curvature (and vicinity) \\
    $\pm$ & spanning regions of negative and positive curvature$^\dag$ \\
    \hline
  \end{tabular*}
\end{table}

It is important to note that in any system with a small number of particles, phase transitions
are somewhat smeared out.  In particular, the onset of ordering in crystal-like states
of our system
is not sharp.  Hence, Table \ref{tab:names} includes a condensed (C) phase within which the
degree of crystallinity varies smoothly from liquid-like to crystal-like.  We quantify
the crystallinity by counting the number of particles $N_{\rm X}$ in a crystalline
environment.
To determine whether a given particle is crystalline, we select all particles within
$r=1.45r_0$ of it.  These $n$ neighbours are gnomonically projected onto the plane
tangent to the surface at the target particle.\cite{Snyder1987A}  The magnitude of the
bond-orientational order parameter is then calculated using
\begin{displaymath}
        |\psi| = \left|\frac{1}{6}\sum_{j=1}^{n}e^{6{\rm i}{\theta}_j}\right|,
\end{displaymath}
where ${\theta}_j$ is the angle between the target particle, the $j^{\rm th}$ neighbour
and some arbitrary local reference direction\cite{Halperin1978A} ($|\psi|$ does not
depend on the choice of this direction).  Particles with $|\psi|>0.6$ are considered
crystalline.  It is important to note that this definition of $|\psi|$ does not count
particles on the edge of an ordered cluster as crystalline because of the missing neighbours.
\par
\begin{figure}[t]
\includegraphics[width=80mm]{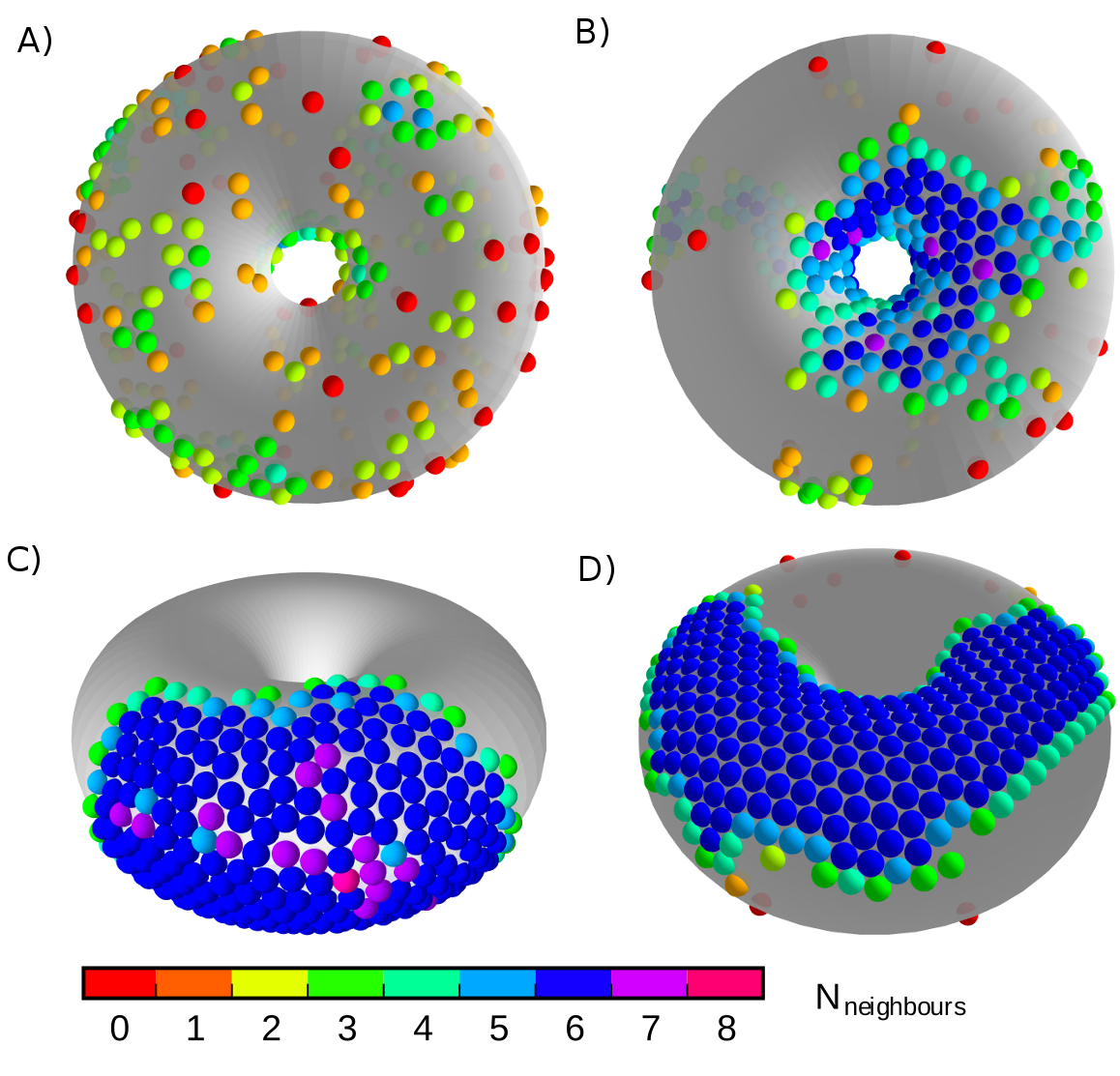}
\caption{\label{fig:torusPic}  Snapshots of the states labelled in the phase diagram of
Fig.~\ref{fig:torusPD}. (A) G, at $\rho=6$, $k_{\rm B}T/\varepsilon=0.72$; (B) L$-$, at
$\rho=6$, $k_{\rm B}T/\varepsilon=0.42$; (C) C+, at $\rho=4$, $k_{\rm B}T/\varepsilon=0.32$;
and (D) X0, at $\rho=18$, $k_{\rm B}T/\varepsilon=0.27$. Particles are coloured by the
number of nearest neighbours.  The states are named according to the convention in
Table~\ref{tab:names}.}
\end{figure}
\begin{figure}[t]
\includegraphics[width=80mm]{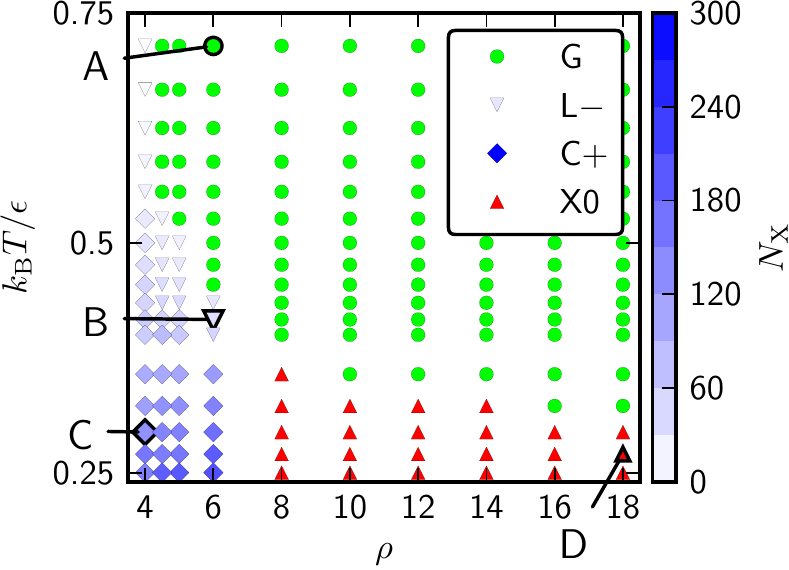}
\caption{\label{fig:torusPD} Phase diagram for 300 Morse particles on a 5-7 torus. The
saturation of the blue symbols represents the crystallinity, showing a steady increase as
the C$+$ state is cooled.  Snapshots of the points labelled A--D are given in
Fig.~\ref{fig:torusPic}.}
\end{figure}

Canonical Monte Carlo simulations of 300 particles on the 5-7 torus reveal
four stable states: G, L$-$, C$+$ and X0.  Snapshots of these states are shown in
Fig.~\ref{fig:torusPic} and their regions of thermodynamic stability are mapped out as
a ``phase diagram'' in the plane of temperature and potential range $\rho$
in Fig.~\ref{fig:torusPD}.  We briefly survey the origin of these states before
examining their coexistence and interconversion in Sec.~\ref{sec:free},
and analysing the competition between them in Sec.~\ref{sec:analysis}.

As expected, the particles cover the torus in a low-density gas-like state G at sufficiently
high temperature for any range of the potential.  Lowering the temperature at long range
($\rho\le6$) leads to a gas--liquid transition.  In the liquid state, the neighbour effect
(described above) is strong enough to drive the cluster to the centre of the torus, where
the mean curvature is largest.  The Gaussian curvature in that region is negative, so this
localised state is denoted L$-$.  Reducing the temperature further produces a driving force
towards crystalline order, but regions of high Gaussian curvature are incompatible with
regular hexagonal packing and the stress effect becomes increasingly important.
The cluster therefore moves to a C$+$ state on the outside of the torus, where the mean
curvature is lower (thereby relieving some stress) but where it can still adopt a compact
shape to reduce its perimeter.  We measure the order in the C$+$ state by the
number $N_{\rm X}$ of crystalline particles, {\it i.e.,} the number of particles with
$|\psi|>0.6$.
Plots and a discussion of the distribution of $|\psi|$ itself can be found in ESI Sec.~S2.
Defined this way, the structure within the C$+$ state varies smoothly between
liquid and crystalline, as indicated by the intensity of the shading in Fig.~\ref{fig:torusPD}
However, the C$+$ state is separated from the neighbouring
L$-$ and X0 states by decisive shifts in location of the cluster.  In Sec.~\ref{sec:free}
we will confirm that the C$+$ state is a distinct free-energy minimum.
\par
Moving at constant temperature to shorter-ranged potentials (horizontally to the
right in Fig.~\ref{fig:torusPD}), another transition is reached.  Deviations from
perfect hexagonal packing become more energetically costly because of the increasing second
derivative at the minimum of the potential well (Fig.~\ref{fig:torusPlan}), and the
stress effect starts to dominate over the perimeter effect.  As a result, a crystal
state, X0, forms as a ribbon on the top (or bottom) of the torus, where the Gaussian
curvature, and therefore the stress, is lowest.  This highly elongated structure comes at
the expense of a long perimeter.  Heating this crystal causes it to sublime directly
back to the G state.
\par
As a reference system without the effects of curvature, we may compare a square plane
of edge $37.17r_0$ with the same area as the 5-7 torus but with conventional periodic
boundary conditions.  The phase diagram for 300 Morse particles on the plane contains
the three standard phases G, L and X and is given in Fig.~S5$^\dag$.  Apart from
the absence of location specifiers, the phase diagram is similar to that in
Fig.~\ref{fig:torusPD}.  The new C$+$ condensed state on the torus mostly occupies
regions that would be crystalline (X) on the plane where the potential is soft enough
for some stress to be accommodated in return for a shorter perimeter.  Relative to the
planar system, curvature also moves the gas--liquid boundary slightly in favour of the
liquid, due to the additional liquid stability that comes from the neighbour effect in
the presence of curvature.
\par
Both the toroidal and planar systems lose their liquid phases at high $\rho$.  This happens
for the same reasons as in three-dimensional systems; a decrease in the range of the
potential causes the gap between the triple point and the gas--liquid critical point
to grow narrower until the gap vanishes altogether, leaving no temperature at which
the liquid is thermodynamically stable.\cite{Hagen1994}
\par
As an example of a torus with a different aspect ratio, we have also
studied the torus with $a=3.5r_0$ and $c=10r_0$, which has the same surface area
as the 5-7 torus.  The thinner tube and wider bore of the 3.5-10 torus introduce
a liquid-like state that is stabilised by wrapping round the tube to reduce the
perimeter.  A more detailed treatment of this case, including the phase diagram,
is provided in ESI Sec.~S3\;D.

\subsection{Free energy surfaces and dynamics\label{sec:free}}

\begin{figure*}[t]
\begin{center}
\includegraphics[width=0.8\textwidth]{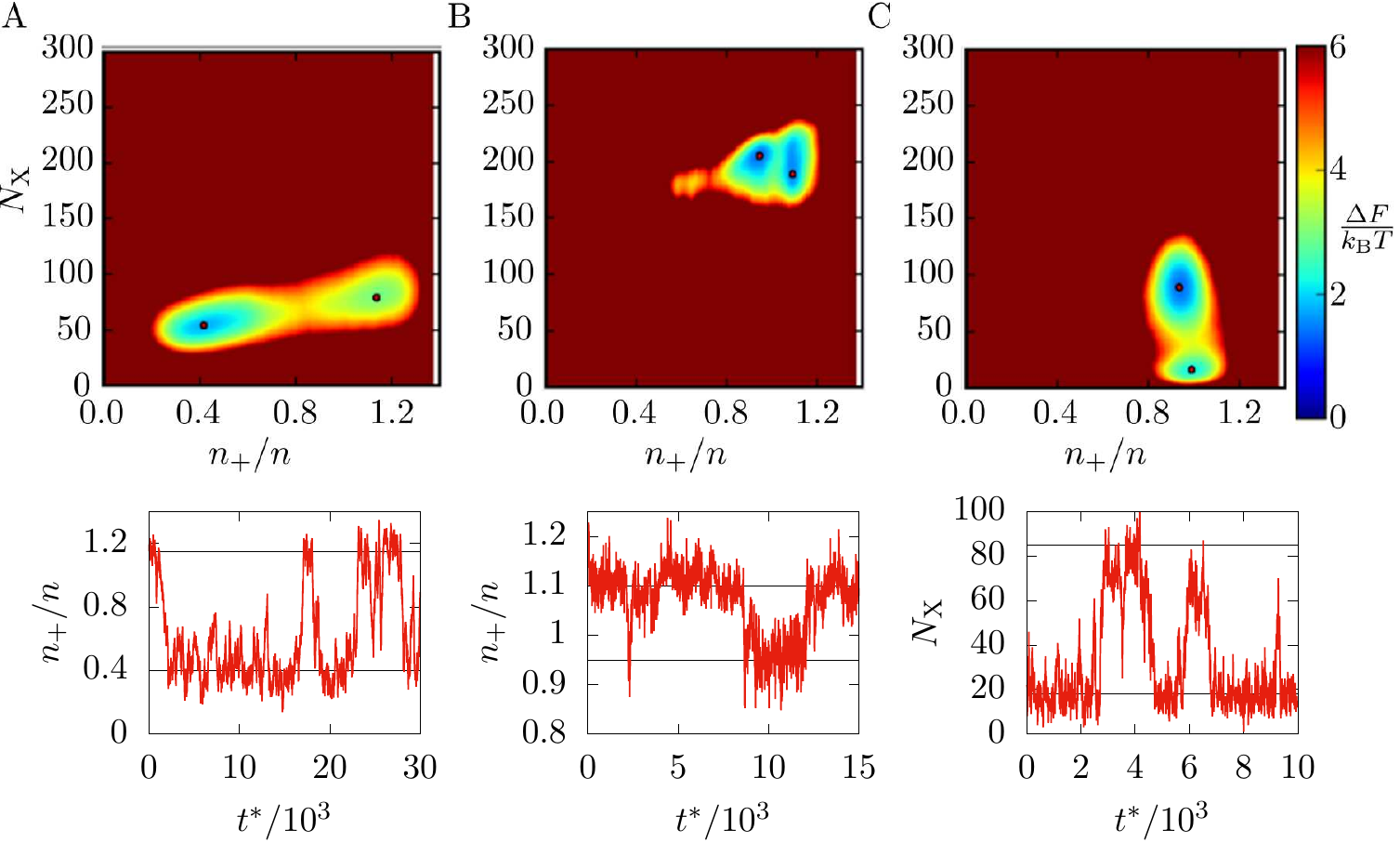}
\caption{\label{fig:heatMaps}
Upper panels: free energy surfaces for 300 Morse particles
on a 5-7 torus with minima highlighted by red dots.
(A) At $\rho=6$ and $k_{\rm B}T/\varepsilon=0.39$.
The red dots indicate the L$-$ (left) and C$+$ (right) states. (B) At $\rho=7$ and
$k_{\rm B}T/\varepsilon=0.25$. The red dots indicate the X0 (left) and C+ (right) states.
(C) At $\rho=20$ and $k_{\rm B}T/\varepsilon=0.30$. The red dots indicate the G (bottom)
and X0 (top) states.
Lower panels: samples of the corresponding
dynamic trajectories switching between the free energy
minima, with horizontal lines denoting the minima.  Time is measured in reduced units
$t^{*} = t\sqrt{\left(\varepsilon/m\right)}/r_{\rm 0}$, where $t$ is the real time and
$m$ is the mass of a particle.
}
\end{center}
\end{figure*}

Returning to the main case of the 5-7 torus,
a direct implication of the phase diagram in Fig.~\ref{fig:torusPD} is that each of the
localised states is the global free energy minimum for a range of $T$ and $\rho$.  At
the boundaries between states, we expect pairs of free energy minima to coexist and to
be separated by a barrier.  Because of the coupling of phase and location, the pathways
between free energy minima must involve migration of matter on the toroidal surface.
\par
To visualise the free energy and trajectories, we project onto the plane of
two order parameters.  To track the phase, one of the order parameters is the number
$N_{\rm X}$ of crystalline particles, as defined in Section \ref{sec:states}.  To track
location, the second order parameter is the density $n_+$ of particles in a region of
positive Gaussian curvature (the ``outside'' of the torus in Fig.~\ref{fig:torusPlan}),
as a fraction of the total density $n$: $n_+ / n$.  The free energy surface is now
constructed by accumulating a two-dimensional histogram $H$ of the order parameters
during a canonical MC simulation and taking $F=-k_{\rm B}T \ln H$.  As we shall see, the
barriers on this surface are sufficiently low that no special sampling techniques are
needed to obtain good statistics.
\par
In Fig.~\ref{fig:heatMaps}, we present the free energy surface at three different points
on the phase diagram.  Each point lies close to a phase transition and shows two minima,
indicating that
two states coexist.  Fig.~\ref{fig:heatMaps}(a) is taken close to the transition between the
L$-$ and C$+$ states.  The transition involves both a dramatic shift of particles towards the
outside of the torus and a slight increase in crystallinity (although at this temperature, the
C$+$ phase is still liquid-like, see Fig.~\ref{fig:torusPD}).
\par
Fig.~\ref{fig:heatMaps}(b) shows the vicinity of the transition from the C$+$ state to the X0
state.  As the temperature here is lower, the C$+$ phase is more crystalline and therefore
appears further along the vertical axis of the free energy plot than in panel (a).  The
evolution of the minimum on the free energy surface
with temperature is smooth and does not involve any barriers,
justifying our treatment of C$+$ as a single phase of intermediate crystalline character.
The figure shows that the transition from C$+$ to X0 is accompanied by a slight increase
in crystallinity and a shift of particles away from the outside of the torus.
\par
The last panel, Fig.~\ref{fig:heatMaps}(c) shows the coexistence between the G and X0 states,
which mainly involves a change in crystallinity as both states are distributed fairly
evenly between the inside and outside of the toroidal surface.
\par
Having identified these coexisting pairs of states, we then used surface-constrained
molecular dynamics simulations to observe the bulk translation of the clusters as the
system switched back and forth between the states.
For each free energy surface in Fig.~\ref{fig:heatMaps}, we
performed a single simulation at fixed temperature and $\rho$ and monitored the order
parameter most relevant to the transition in question.  The plot below
each free energy map shows
part of the resulting equilibrium time trace of the order parameter.  In each case, the
trace shows rapid switches of the system between two states, with comparatively long
residence times within the states, confirming the interpretation of the distinct
macrostates identified in Sec.~\ref{sec:states}.  In the ESI$^\dag$ we have
included videos that visualise an example pathway for each of the three transitions.

\subsection{The competing effects of curvature\label{sec:analysis}}

In this section, we provide a more quantitative analysis of the three effects
of curvature identified in Sec.~\ref{sec:states}.  The results provide insight
into the competition between the effects and the factors that influence the
points where they balance.
In turn, this helps to generalise the principles to other examples of non-uniformly
curved two-dimensional systems.
\par
\begin{figure}[t]
\includegraphics[width=90mm]{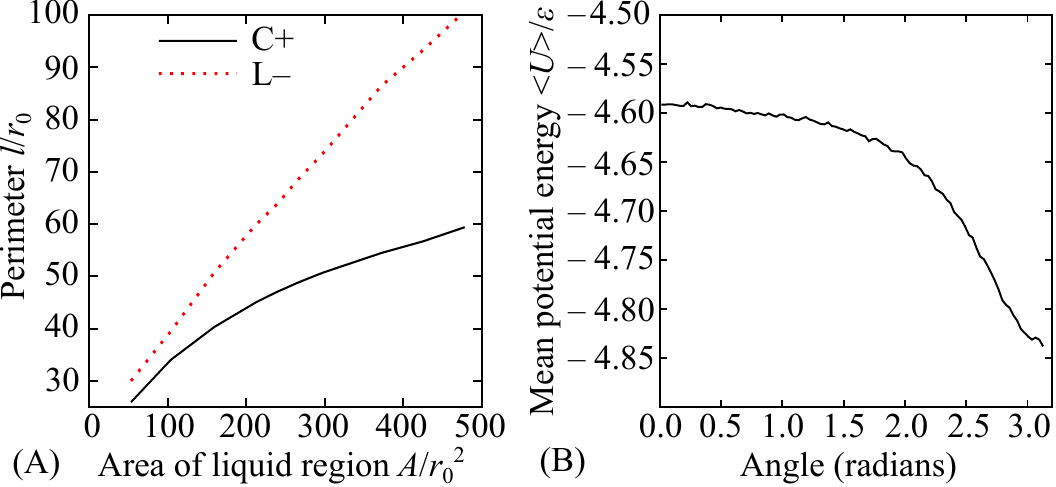}
\caption{\label{fig:angleEn}
(A) Minimised perimeter of the liquid--gas interface for the C$+$ and L$-$
states as a function of the area of the liquid region.
(B) Average potential energy per particle in a
surface-covering liquid state as a function of the poloidal angle $\theta$,
taken from a grand canonical simulation at $\rho=6$,
$k_{\rm B}T/\varepsilon = 0.48$ and activity $z = 57.544$.}
\end{figure}

For liquid-like states, the optimal location is determined by the perimeter and
neighbour effects.  On the torus, these two effects compete with each other because
the most strongly curved
region (beneficial for next-nearest neighbours) is on the inside of the torus, but the
sign of the Gaussian curvature is negative there, so the perimeter is larger for a given
area (costing line energy).  We have obtained the optimal perimeters as a function of
the enclosed area using constrained minimisation in the Surface Evolver software.\cite{brakke1992A}
By carefully choosing the symmetry of the initial patch about poloidal angles $\theta=0$
and $180^\circ$, the optimisation can be performed separately for patches in the regions
of positive and negative curvature, respectively.  The two perimeters are plotted in
Fig.~\ref{fig:angleEn}(A), showing the increasing advantage of the C$+$ state with area.
\par
The contribution of the potential energy to the neighbour effect can also be
quantified by examining the mean energy per particle as a function of the poloidal
angle $\theta$ in a liquid that covers the whole surface.  We have obtained a
specimen system-covering fluid by performing a grand canonical simulation of particles
with range parameter $\rho=6$ at a temperature of $k_{\rm B}T/\varepsilon=0.48$ and
activity $z=57.544$.  The location dependence of the mean potential energy is shown
in Fig.~\ref{fig:angleEn}(B), where it can be seen that the energy per particle is
some $5\%$ lower on the inside of the torus than on the outside.

\begin{figure}[t]
\begin{center}
\includegraphics[width=70mm]{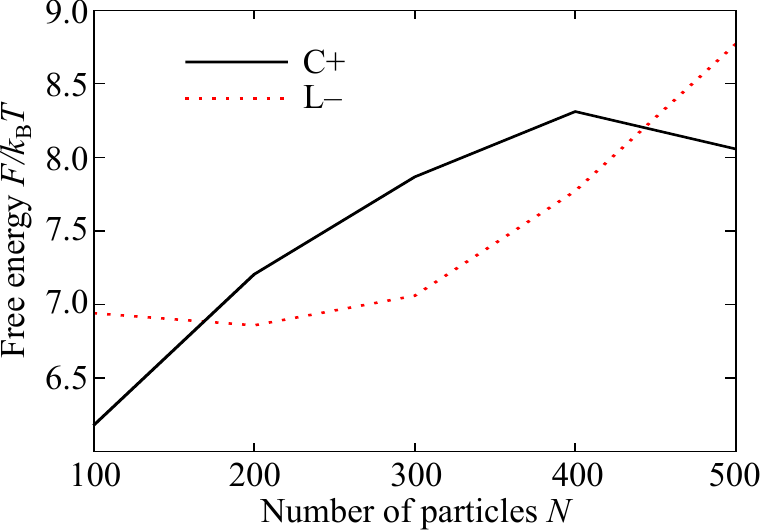}
\end{center}
\caption{\label{fig:liqFree} Depth of the free energy minima of the L$-$ and C$+$ states on
the 5-7 torus for $\rho=6$ at reduced temperature $k_{\rm B}T/\varepsilon = 0.37$.
The origin of the vertical scale is arbitrary; only comparisons between the two lines
matter.}
\end{figure}

The plots in Fig.~\ref{fig:angleEn} demonstrate two of the key ingredients in
the perimeter and the neighbour effects.  Other important considerations include
the line tension on the perimeter itself, which is much harder to evaluate.
However, the full complexity of the resulting
competition between the C$+$ and L$-$ states can be seen by tracking the
relative depths of the corresponding free energy minima in Fig.~\ref{fig:heatMaps}(A)
as a function of the total number $N$ of particles in the system.  For potential
range $\rho=6$ and temperature $k_{\rm B}T/\varepsilon=0.37$, the
two states coexist as (meta)stable minima at least from $N=100$ to $300$
and their relative free energies are shown over this range in Fig.~\ref{fig:liqFree}.
Interestingly, the lines cross twice, emphasising the nonlinearity of the perimeter
and neighbour effects with respect to $N$.

\begin{figure}[t]
\begin{center}
\includegraphics[width=90mm]{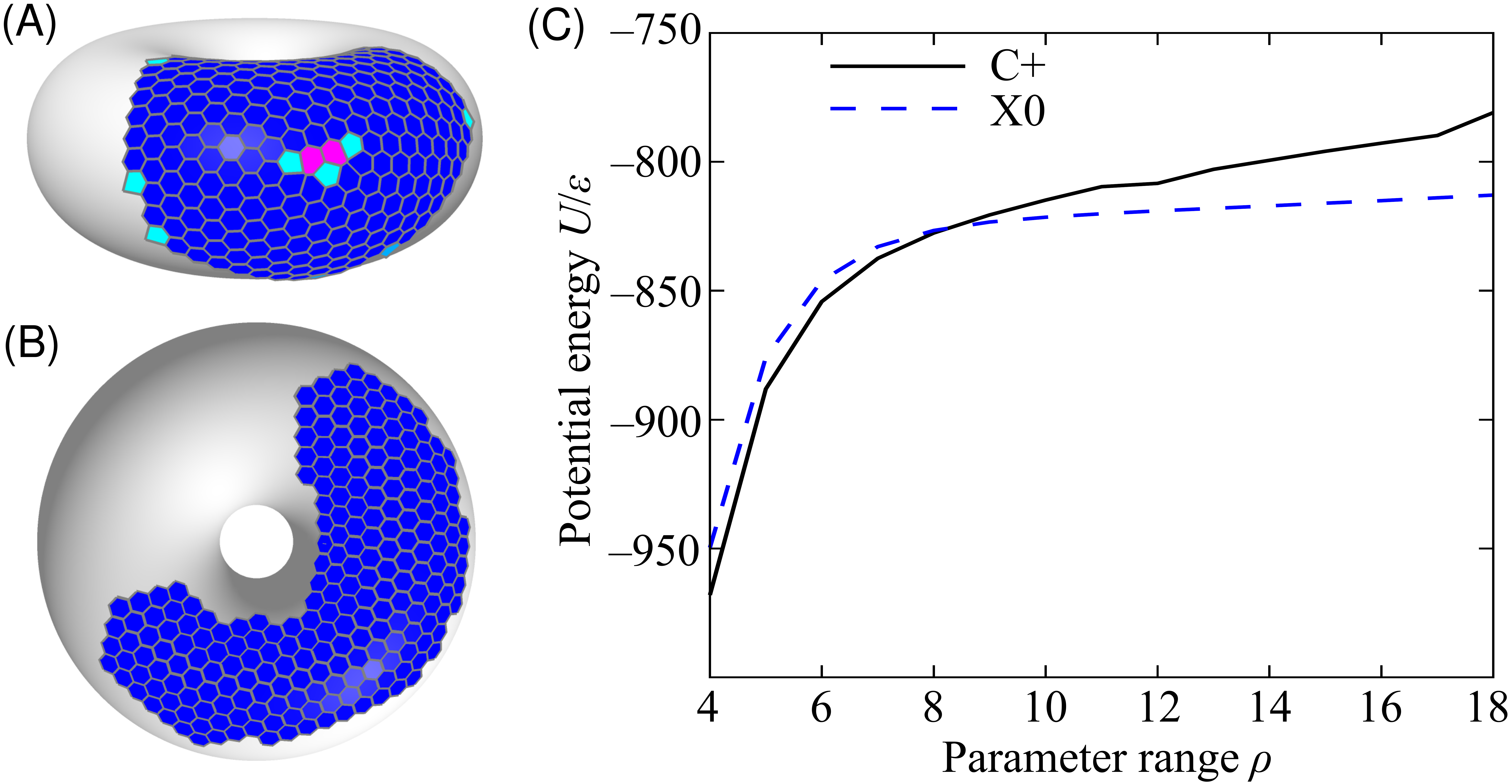}
\caption{\label{fig:crystalTot}
Putative global potential energy minimum structures
for $N=300$ particles on the 5-7 torus for (A) range parameter $\rho=4$ (C+ configuration),
(B) $\rho=18$ (X0).  The structures are depicted by their Voronoi tessellations and
the colour of each cell corresponds to the coordination number of the particle that it contains
(colour scheme as in Fig.~\ref{fig:torusPic}).
(C) Potential energy of the C+ and X0 minima as a
function of the range parameter $\rho$ for $N=300$ particles on the 5-7 torus.}
\end{center}
\end{figure}

The phase diagram in Fig.~\ref{fig:torusPD} shows that there is
a competition between the C$+$ and X0 states as a function of the potential range
parameter $\rho$ at low temperature.  Under these conditions,
the potential energy is the dominant contribution to the free energy and it is
instructive to locate the global potential energy minima using BHPT.
The optimisation runs used eight parallel replicas with an exponential distribution
of temperatures in the range $0.3\le k_{\rm B}T/\varepsilon \le 2$ and up to 60,000
basin-hopping steps per case.  Runs were initiated with quenched structures from
canonical MC simulations.
Fig.~\ref{fig:crystalTot} shows the resulting putative global minima for
a soft ($\rho=4$) and a stiff ($\rho=18$) interaction potential for our case study
of $N=300$, while sequences of global minima from $N=100$ to $500$ are
presented in Sec.~S1\;A of the ESI$^\dag$.  To highlight
any packing defects in these structures, we have depicted them by their Voronoi
tesselations.  To avoid ill-defined Voronoi cells at the edges of the clusters,
any Voronoi vertex lying further than $1.3r_0$ from its particle has been deleted
in the analysis, and any Voronoi cell with fewer than five edges is not displayed
in the figure.  These measures have the effect of discarding most cells corresponding
to edge particles but they do not alter the depiction of the
cluster interiors.
\par
The long-ranged potential is able to accommodate the stress on the hexagonal
lattice in the region of positive Gaussian curvature on the outside of the torus,
characteristic of the C$+$ thermodynamic state [Fig.~\ref{fig:crystalTot}(A)].
This stress is manifest mostly as a systematic and delocalised distortion of
the packing, but a packing defect is also visible in the centre of the cluster.
The defect consists of three pentagonal tiles and two heptagonal tiles, giving
a positive overall topological charge of $+1$, as expected in a region of
positive Gaussian curvature.\cite{Burke15a}
Such defects are too costly for short-ranged potentials because a lower formal
coordination number involves losing most (not just some) of the interaction with
a neighbouring particle.  Even delocalised stress is highly unfavourable because of the
sensitivity of the potential near its minimum.  Hence, the global minimum
changes to lie in the flattest region of the surface like the X0 thermodynamic
state [Fig.~\ref{fig:crystalTot}(B)].  The cluster is now far less compact and the
increase in its perimeter, demarcated by low-coordination particles, is readily visible.
\par
We can estimate the point at which the X0 state takes over from C$+$ by tracking the
energy of the two structures depicted in Fig.~\ref{fig:crystalTot} as a function of
$\rho$.  We do this by incremental changes in $\rho$, each followed by a local
minimisation to relax the structure to the point of mechanical equilibrium without
significant rearrangement.  Both structures persist as (meta)stable minima over
a wide range of $\rho$.  The resulting potential energy curves are compared in
Fig.~\ref{fig:crystalTot}(C), showing that they cross at $\rho\approx8$, which is
consistent with the boundary between C$+$ and X0 in the finite-temperature phase
diagram of Fig.~\ref{fig:torusPD}.  Further insight can be gained from
the comparison in Fig.~\ref{fig:crystalTot}(C) by decomposing the energy into
contributions with a direct physical interpretation.  This analysis is provided in
Sec.~S1\;B of the ESI$^\dag$.

The competition between the perimeter, stress and
neighbour effects is altered by the scale, as well as by the shape,
of the surface.  Any given torus has a finite
surface area and --- like a sphere --- can only accommodate a limited number of particles before
overcrowding incurs a steep free energy penalty.\cite{Law2018A}  However, consider
a thought experiment in which the two radii $a$ and $c$ of the torus
are both increased by a factor $f$,
keeping the aspect ratio fixed, and the number of particles $N$ is increased by $f^2$ to
keep the surface coverage approximately equal.  This scaling would uniformly reduce the
Gaussian curvature by a factor of $1/f^2$.  Hence, the neighbour effect, which relies
on local curvature, would become less important with increasing $f$, and this is likely
to destabilise the L$-$ state.  However, the perimeter
would increase approximately as $f$ (for a given cluster shape) and,
for states with crystalline character, there would also be an increase in stress.\cite{Meng2014A}
As we have seen (for example, in Fig.~\ref{fig:crystalTot}), stress introduces a complex interplay
between elastic energy and defects, the energetic cost of which depends on the interaction potential.
Hence, we can expect a non-trivial evolution of structure and stability with overall scale
of the host surface, even at fixed aspect ratio.

\subsection{A sinusoidal surface}

\begin{figure*}
\begin{center}
\includegraphics[width=0.9\textwidth]{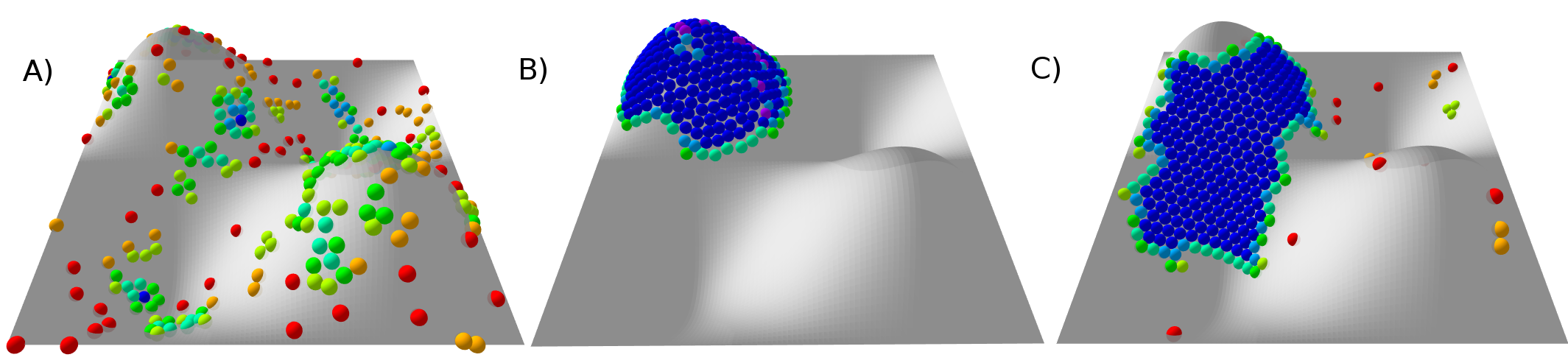}
\caption{\label{fig:eggboxPic}
Snapshots of states for 300 particles on a periodic sinusoidal surface. (A) G, at $\rho=4$,
$k_{\rm B}T/\varepsilon=0.78$; (B) C+, at $\rho=4$, $k_{\rm B}T/\varepsilon=0.39$; and (C)
X0, at $\rho=20$, $k_{\rm B}T/\varepsilon=0.27$.  Particles are coloured by the number of
nearest neighbours (see Fig.~\ref{fig:torusPic} for key).}
\end{center}
\end{figure*}

To demonstrate that the phenomena seen on the torus can be extended to other curved surfaces,
we have performed simulations on a sinusoidal surface with periodic boundaries, defined in
Cartesian coordinates $x,y,z$ by
\begin{displaymath}
z = h\sin\left(2\pi x/L\right)\sin\left(2\pi y/L\right).
\end{displaymath}
We set $h=7.5r_0$ and $L=30.75r_0$ so that the area and maximum Gaussian curvature of this
surface match those of the 5-7 torus. The metric $g$ factors for
use in Eq.~(\ref{eq:acceptance}) are now given by
\begin{displaymath}
g = \sqrt{\left(\frac{2\pi h}{L}\right)^2
\left[1-\cos\left(\frac{4 \pi x}{L}\right)\cos\left(\frac{4 \pi y}{L}\right)\right]+1}.
\end{displaymath}

\begin{figure}[t]
\includegraphics[width=80mm]{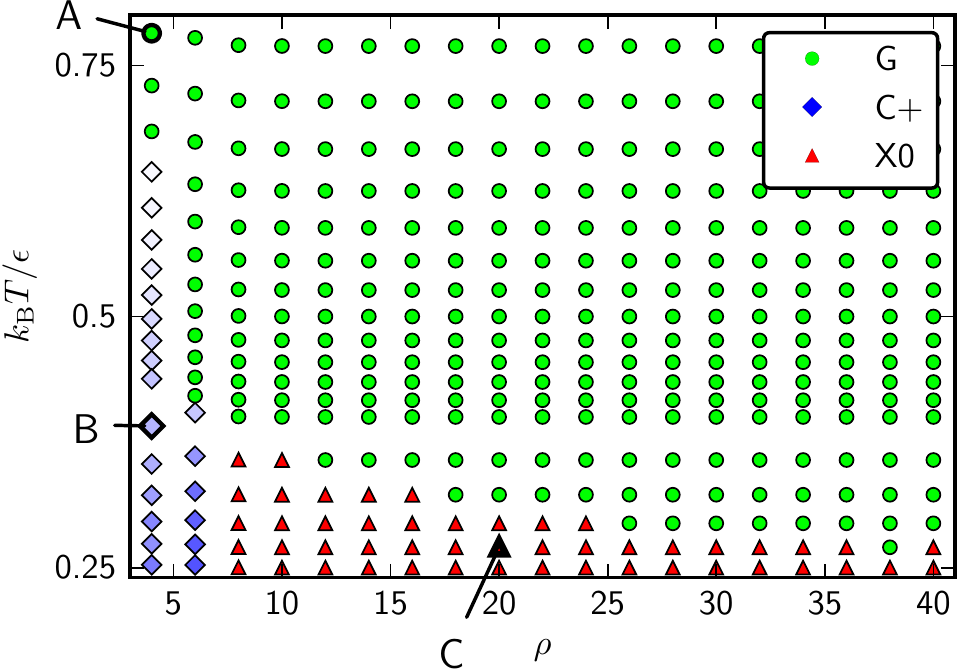}
\caption{\label{fig:eggPD} Phase diagram for 300 Morse particles on a periodic sinusoidal
surface as a function of the potential range parameter $\rho$ and the reduced temperature
$k_{\rm B}T/\varepsilon$.  Snapshots of the points labeled A--C are given in
Fig.~\ref{fig:eggboxPic}.}
\end{figure}

Snapshots of the stable states of this system are presented in Fig.~\ref{fig:eggboxPic}
and the phase diagram is shown in Fig.~\ref{fig:eggPD}.  This system has only three states:
G, C+ and X0.  At high temperature, the system is found in the G state.  For softer
(longer-ranged) potentials, as the system is cooled the particles condense into a liquid
phase, C$+$, which is always found on one of the peaks or troughs of the surface.  This
configuration is preferred both for its short perimeter and for its high mean curvature
(leading to a favourable neighbour effect).  As the system is cooled further, the condensed
cluster becomes more crystalline.  Although the contribution of stress in the crystal to
the free energy is increasing, at low $\rho$ the cluster does not move again, unlike on
the torus which has the transition from L$-$ to C$+$.  We suggest that this is because
the line tension is high enough that, unlike on the torus, moving to a less curved region
with a longer perimeter would not be favourable.  However, at higher values of $\rho$, the system
crystallises around the flanks of the peaks, where the Gaussian curvature is lowest, giving
an X0 state.  As in the corresponding X0 state on the torus, as well as in the branched
structures observed by Meng {\it et al.}\cite{Meng2014A} on spherical droplets, a longer
perimeter is traded for less frustration in stiffer crystals.

An important message from comparing the torus and the sinusoidal surface is that the
relationship between the perimeter, stress and neighbour effects is determined by the
shape of the surface.  For example, the regions of largest principal curvature on the sinusoidal
surface are the peaks and troughs, where the Gaussian curvature is positive.  Hence,
the perimeter and neighbour effects reinforce each other in this region.  In contrast,
the largest principal curvature on a torus is around the central bore, where the
Gaussian curvature is negative, leading to antagonism between the perimeter and
neighbour effects.  Nevertheless, it is the same set of physical arguments that
comes into play in all cases.

\section{Conclusions}

We have shown that, in non-uniformly curved two-dimensional systems, clusters of attractive
colloids minimise their free energy by adopting specific shapes and translating to specific
locations.  The equilibrium shape and location depends not only on the phase of matter in
the cluster,
but also on the range of the interaction potential and the curvature of the underlying
surface.  The coupling of phase to shape and location leads to dramatic reorganisation of
matter as the conditions are varied.  In particular, phase transitions can be accompanied
by wholesale migration of matter to different parts of the surface.  We have demonstrated
that these effects arise in systems where the particles themselves are simple spheres with
isotropic interactions.  These particles collectively respond to curvature despite having
no individual preference for a particular curvature or a curvature-adapted shape.
\par
We have identified three universal contributions to the free energy that drive the
behaviour: the length of the cluster perimeter, the stress on packing induced by
Gaussian curvature, and the distances from a given particle to its next-nearest
neighbours.  These considerations explain the four phase--location coupled states
that we observe on a torus and the three states on a sinusoidal surface.  Free energy
calculations also showed the barriers between these states and molecular dynamics
simulations confirmed the switching between them.
\par
There are a number of avenues for future investigations in which the
phase behaviour is strongly affected by non-uniform curvature.  For instance, we expect
additional levels of control and rich behaviour in cases where surface curvature
is coupled with an anisotropic interaction potential or polydisperse mixtures.
It would also be interesting to analyse in detail the defects observed in
the C$+$ and X0 states, at both zero and finite temperature, and in particular to study
whether they follow existing predictions on how the number of defects depends on the amount
of curvature enclosed\cite{Burke15a,Li19c} and the effects of thermal
fluctuations.\cite{Paquay16b}  Another open question is the case where the confining surface is
flexible, where the curvature responds to the particles that are confined upon it.  This latter
form of coupling is relevant, for example, in a variety of clustering and aggregation
phenomena on lipid membranes.\cite{Vahid2017A}
\par
We hope our work will motivate experimental demonstrations of the cooperative
curvature-sensing effects shown in this article, in both biological and engineered
systems.  Due to the generality of the effects, the non-uniformly curved surface need
not be specifically toroidal or sinusoidal as in the examples presented here.

\section*{Conflicts of interest}
There are no conflicts to declare.

\section*{Acknowledgements}
The authors thank Stefan Paquay for advice on the MD simulations.
J.O.L.~is grateful to the Engineering and Physical Sciences
Research Council (EPSRC) Centre for Doctoral Training in Soft Matter and Functional Interfaces
(SOFI, grant EP/L015536/1) for financial support.

%%%END OF MAIN TEXT%%%

%The \balance command can be used to balance the columns on the final page if desired. It should be placed anywhere within the first column of the last page.

\balance

%If notes are included in your references you can change the title from 'References' to 'Notes and references' using the following command:
%\renewcommand\refname{Notes and references}

%%%REFERENCES%%%
%\bibliographystyle{rsc}
%\bibliography{refs}

\providecommand*{\mcitethebibliography}{\thebibliography}
\csname @ifundefined\endcsname{endmcitethebibliography}
{\let\endmcitethebibliography\endthebibliography}{}

\end{document}